\documentclass[a4paper,12pt]{article}
\usepackage[cp1251]{inputenc}
\usepackage[english,russian]{babel}
\usepackage{amsmath}
\usepackage{amssymb}
\usepackage{amsfonts}

\usepackage{hyperref}
\usepackage{cmap}

\def\eq#1{\begin{equation}#1\end{equation}}
\def\eqs#1{\begin{eqnarray}#1\end{eqnarray}}
\def\eqn#1{\begin{equation}\begin{split}#1\end{split}\end{equation}}
\def\seq#1{\begin{equation*}#1\end{equation*}}

\def\bN{\beta_N}
\def\Z{\mathbb Z}
\def\N{\mathbb N}

\def\wt{\widetilde w}

\title{\bf Модифицированные серии интегрируемых дискретных уравнений на квадратной решетке с нестандартной симметрийной структурой}% ????????? ???????? ??????
\author{\bf Р.Н. Гарифуллин$^{ab}$, Р.И. Ямилов$^a$
\\ $^a$ Институт математики с ВЦ УФИЦ РАН,\\ 450008, г. Уфа, ул.Чернышевского, 112, Россия
\\ $^b$ Башкирский государственный университет,\\ 450076, г. Уфа, ул. Заки Валиди, 32, Россия\\
\sl E-mails: rustem@matem.anrb.ru, RvlYamilov@matem.anrb.ru}%\\{\sl URL: \url{http://matem.anrb.ru/garifullinrn},} \\ {\sl\url{http://matem.anrb.ru/en/yamilovri}}}
\date{}

\begin{document}
\begin{center}
\selectlanguage{russian}
{\large \bf Modified series of integrable discrete equations on a square lattice with a non-standard symmetry structure}\\
\vspace{0.5cm}
{\bf R.N. Garifullin$^{ab}$ and R.I. Yamilov$^a$}
\vspace{0.5cm}
\\ $^a$ Institute of Mathematics, Ufa Federal Research Centre RAS,\\ 112 Chernyshevsky Street, Ufa 450008, Russia
\\ $^b$ Bashkir State University, 32 Zaki Validi Street, Ufa 450074, Russia\\
\vspace{0.3cm}
\sl E-mails: rustem@matem.anrb.ru, RvlYamilov@matem.anrb.ru %\\{\sl URL: \url{http://matem.anrb.ru/garifullinrn},} \\ 
\end{center}
\vspace{1.0cm}

{\bf Abstract. }{ In a recent paper [TMP, {\bf 200}: 1 (2019), 966--984] by the authors, a series of integrable discrete autonomous equations on a square lattice with a non-standard structure of generalized symmetries is constructed. We build modified series by using discrete non-point transformations. We use both non-invertible linearizable transformations and non-point transformations invertible on solutions of the discrete equation.

As a result, we get several series of new examples of discrete equations along with their generalized symmetries and master symmetries. The generalized symmetries constructed give new integrable examples of five- and seven-point differential-difference equations together with their master symmetries. In the case of discrete equations, the method of constructing non-invertible linearizable transformations by using conservation laws is considered, apparently, for the first time.
}


\newpage
\selectlanguage{russian}
\maketitle

\abstract{В недавней работе авторов [ТМФ, {\bf 200}:1 (2019), 50--71] построена серия интегрируемых дискретных автономных уравнений на квадратной решетке с нестандартной структурой высших симметрий. Мы строим модифицированные серии, используя дискретные неточечные преобразования. Мы используем как необратимые линеаризуемые преобразования, так и неточечные преобразования, обратимые на решениях дискретного уравнения.

В результате мы получаем серии новых примеров дискретных уравнений вместе с их высшими симметриями и мастер-симметриями. Построенные высшие симметрии дают новые интегрируемые примеры пяти- и семиточечных дифференциально-разностных уравнений вместе с их мастер-симметриями. В случае дискретных уравнений, метод построения необратимых линеаризуемых преобразований при помощи законов сохранения рассматривается, по-видимому, впервые.}

\medskip
{\bf Ключевые слова: }дискретное уравнение, высшая симметрия, мастер-симметрия, дифференциально-разностное уравнение, необратимое преобразование.

\section{Введение}
В работе \cite{gy19} представлена бесконечная серия дискретных уравнений вида
\eqn{(u_{n,m+1}+1)(u_{n,m}-1)=\beta_N(u_{n+1,m+1}-1)(u_{n+1,m}+1),\label{our_bN}} где $n,m\in\Z$ -- дискретные независимые переменные, $u_{n,m}$ -- неизвестная функция двух дискретных переменных. Уравнения серии нумеруется натуральным числом $N$, а коэффициенты $\beta_N$ являются корнями из единицы: $\beta_N^N=1,\  N\geq1$.

Чтобы разделять уравнения с разными $N$, мы рассматриваем примитивные корни из единицы. Ясно, что $\beta_1=1$.
При $N>1$ примитивные корни определяются следующим образом:
\eq{\label{bN} \beta_N^N=1,\qquad \beta_N^j\neq 1, \ \   1\leq j<N. }  В частности, \eqn{\beta_1=1,\quad \beta_2= -1,\quad \beta_3=-\frac12\pm i\frac{\sqrt3} 2,\quad \beta_4=\pm i,\label{our_b1234}} т.е. в двух последних случаях имеется два примитивных корня, соответствующих знакам $+$ и $-$. Для любого $N>2$ существует как минимум два примитивных корня: $\beta_N=\exp(\pm 2i\pi/N)$. Таким образом, мы рассматриваем серию уравнений вида \eqref{our_bN},  в которых коэффициенты $\beta_N$ являются примитивными корнями $N$-ой степени из единицы.

Уравнения серии \eqref{our_bN} являются интегрируемыми в том смысле, что они имеют бесконечные иерархии высших симметрий и законов сохранения так же, как и $L-A$ пары. Среди этих симметрий и законов сохранения можно выделить бесконечные иерархии автономных.

Аналогичные серии неавтономных дискретных уравнений рассматривались в \cite{ghy15}. Серии дискретных уравнений, интегрируемых по Дарбу, и уравнений типа Бюргерса изучались в \cite{gy12u,gy19u}.   
 
При помощи неточечных преобразований мы будем получать в этой статье новые серии дискретных уравнений вместе с высшими симметриями. Для простоты мы ограничимся несколькими простейшими высшими симметриями в каждом из направлений.

Как показано в работе \cite{gy19}, высшие симметрии первого и второго порядка в $m$-направлении для любого уравнения серии \eqref{our_bN} имеют вид:
\eq{\partial_{t_1} u_{n,m}=\beta_N^n (u_{n,m}^2-1)(u_{n,m+1}-u_{n,m-1}),\label{mV1}} 
\eqn{\partial_{t_2}u_{n,m}&=\beta_N^{2n}(u_{n,m}^2-1)\big[(u_{n,m+1}^2-1)(u_{n,m+2}+u_{n,m})\\&-(u_{n,m-1}^2-1)(u_{n,m}+u_{n,m-2})-4(u_{n,m+1}-u_{n,m-1})\big].\label{sym_mN2}}
В каждом порядке имеется аналогичная высшая симметрия в $m$-направлении. Структура этих симметрий  такова, что уравнение \eqref{our_bN} с номером $N$ обладает автономными высшими симметриями порядков $kN$, $k\in \N$.

В $n$-направлении структура высших симметрий нестандартна. Вид и порядок простейшей симметрии в этом направлении зависит от числа $N$.
Простейшая высшая симметрия уравнения серии \eqref{our_bN} с $N=1$ имеет первый порядок: 
\eq{\partial_{\theta_1} u_{n,m}=(u_{n,m}^2-1)\left(\frac{a_{n+1}}{u_{n+1,m}+u_{n,m}}-\frac{a_{n}}{u_{n,m}+u_{n-1,m}}\right),\label{sym_nN1}}
где $a_n=b+cn$, а $b, c$ -- произвольные постоянные. 

При $N=2$ простейшая высшая симметрия имеет второй порядок: 
\eqn{\partial_{\theta_2} u_{n,m}=&(u_{n,m}^2-1)(T_n-1)\left(\frac{a_{n+1}(u_{n+1,m}+u_{n,m})}{U_{n,m}}+\frac{a_{n}(u_{n-1,m}+u_{n-2,m})}{U_{n-1,m}}\right),\label{sym_nN2}\\ U_{n,m}=&(u_{n+1,m}+u_{n,m})(u_{n,m}+u_{n-1,m})-2(u_{n,m}^2-1).}
 Здесь функция $a_n$ имеет вид $a_n=b_n+cn$, где $c$ -- константа и $b_{n}\equiv b_{n+2}$ -- произвольная двухпериодическая функция от $n$. Ее можно представить в виде $b_n=b^{(1)}+b^{(2)}(-1)^n$ с двумя произвольными постоянными $b^{(1)},\ b^{(2)}$. Оператор сдвига вдоль направления $n$ обозначен здесь и далее через $T_n$: $T_n h_{n,m}=h_{n+1,m}.$

Простейшая высшая симметрия уравнения \eqref{our_bN} c $N=3$ имеет третий порядок:
\eqn{\partial_{\theta_3} u_{n,m}=&(u_{n,m}^2-1)(T_n-1)\left(\frac{a_{n+2}V_{n,m}}{U_{n,m}}+\frac{a_{n}W_{n,m}}{U_{n-2,m}}+(T_n+1)\frac{a_{n+1}Z_{n,m}}{U_{n-1,m}}\right)\label{sym_nN3},\\V_{n,m}=&\beta_3^2(u_{n+1,m}^2-1)+u_{n+1,m}(u_{n+2,m}-u_{n-1,m})-u_{n+2,m}u_{n-1,m}+1,\\W_{n,m}=&\beta_3(u_{n-2,m}^2-1)+u_{n-2,m}(u_{n-1,m}+u_{n-3,m})+u_{n-1,m}u_{n-3,m}+1,\\Z_{n,m}=&(u_{n+1,m}+u_{n,m})(u_{n-1,m}+u_{n-2,m}),\\U_{n,m}=&\beta_3^2(u_{n+1,m}^2-1)(u_{n,m}+u_{n-1,m})+\beta_3(u_{n,m}^2-1)(u_{n+2,m}+u_{n+1,m})\\+(&u_{n+1,m}u_{n,m}+1)(u_{n+2,m}+u_{n-1,m})+(u_{n+1,m}+u_{n,m})(u_{n+2,m}u_{n-1,m}+1).} 
Здесь $\beta_3$ -- любой из двух примитивных корней из единицы, показанных в \eqref{our_b1234}. Функция $a_n$ задается формулой $a_n=b_n+cn$, где $c$ -- константа, а $b_{n}\equiv b_{n+3}$ -- произвольная трехпериодическая функция. Она может быть представлена в виде $b_n=b^{(1)}+b^{(2)}\beta_3^n+b^{(3)}\beta_3^{2n}$, где $b^{(1)},\ b^{(2)},\  b^{(3)}$ -- произвольные постоянные. 

Каждая из приведенных выше формул \eqref{sym_nN1}-\eqref{sym_nN3} содержит автономную симметрию, соответствующую $a_n\equiv 1$.
Случай $a_n=n$ играет роль мастер-симметрии для соответствующего дискретного уравнения \eqref{our_bN}. Формулы \eqref{sym_nN1}-\eqref{sym_nN3} при любом фиксированном $m$ определяют интегрируемые дифференциально-разностные уравнения при $c=0$ и их мастер-симметрии при $a_n=n$. 

При помощи неточечных преобразований мы получаем в этой работе из формул \eqref{sym_nN1}-\eqref{sym_nN3} новые примеры автономных пяти- и семиточечных дифференциально-разностных уравнений вместе с их мастер-симметриями.

Неточечные преобразования, которые мы используем для построения модифицированных серий и высших симметрий делятся на два типа. К первому типу относятся дискретные необратимые линеаризуемые преобразования, введенные в работе \cite{gyl16}, которые строятся при помощи специальных законов сохранения. В случае дискретных уравнений, используемый здесь метод построения линеаризуемых преобразований при помощи законов сохранения является, по-видимому, новым.

Ко второму типу относятся неточечные преобразования, обратимые на решениях дискретного уравнения \cite{s10,y91}. В случае высших симметрий и соответствующих им дифференциально-разностных уравнений такие преобразования приводят к необратимым преобразованиям типа Миуры или сложно-линеаризуемым преобразованиям.

В секции \ref{seclin} мы рассматриваем дискретные необратимые линеаризуемые преобразования. Сначала мы излагаем процедуру построения таких преобразований при помощи специальных законов сохранения и схему построения модифицированных дискретных уравнений и соответствующих высших симметрий. Затем, в результате применения этого метода мы получаем четыре модифицированные серии дискретных уравнений вместе с простейшими высшими симметриями. В секции \ref{secobr} мы используем неточечные преобразования, обратимые на решениях дискретного уравнения. Это позволяет нам построить еще две модификации вместе с высшими симметриями и мастер-симметриями.

\section{Линеаризуемые преобразования}\label{seclin}
Следуя работе \cite{gyl16}, где обсуждаются дифференциально-разностные уравнения, мы вводим здесь понятие линеаризуемого преобразования. Мы будем рассматривать в этой секции преобразования первого порядка, которые линеаризуются при помощи точечных преобразований. Такое  преобразование в $n$-направлении может быть записано в виде:
\eq{U_{n,m}=F_{n,m}[(T_n-1)G_{n,m}(V_{n,m})],\label{lin_tr}}
где $F'_{n,m}(x)\neq 0$ и $G'_{n,m}(x)\neq 0$ для любых $n,m$. Используя очевидные неавтономные точечные замены $U_{n,m}$ и $V_{n,m}$, мы можем привести такое преобразование к виду: \seq{\widehat U_{n,m}=(T_n-1)\widehat V_{n,m},} который представляет собой простейшее линейное преобразование. 

Композиции таких преобразований также называются линеаризуемыми \cite{gyl16}. В случае, когда композиция состоит из преобразований разных направлений, мы получаем сложно-линеаризумое преобразование, см. пример \eqref{vu2} в секции \ref{sec_first}. 
Заметим, что преобразования типа Миуры к линеаризуемым не относятся \cite{gyl16}. Пример такого преобразования представлен в секции \ref{sec_sec}, см. \eqref{wv2}.

В $m$-направлении, линеаризуемые преобразования первого порядка, рассматриваемые в этой секции, имеют аналогичный вид:
\seq{U_{n,m}=F_{n,m}[(T_m-1)G_{n,m}(V_{n,m})],} где  через $T_m$ обозначен оператор сдвига вдоль направления $m$: $T_m h_{n,m}=h_{n,m+1}.$

\subsection{Построение простейших линеаризуемых преобразований}\label{sec_th}
Предположим, что уравнение \eqref{our_bN} имеет закон сохранения вида:
\eq{\label{zsoh}(T_m-1)p_{n,m}(u_{n,m})=(T_n-1)q_{n,m}(u_{n,m+1},u_{n,m}),} где $p'_{n,m}(x)\neq 0$ для любых $n,m$, т.е.  функция $p_{n,m}(x)$ обратима при всех $n,m$.  Можно ввести новую неизвестную функцию $v_{n,m}$: 
\eq{\label{puv}p_{n,m}(u_{n,m})=(T_n-1)r_{n,m}(v_{n,m}),}где $r'_{n,m}(x)\neq 0$ для любых $n,m$. Это соотношение дает явную формулу для $u_{n,m}$:\eq{\label{uv}u_{n,m}=p_{n,m}^{-1}[(T_n-1)r_{n,m}(v_{n,m})].} 

Подставляя в \eqref{zsoh} функцию \eqref{uv} и интегрируя  по дискретной переменной $n$, т.е. применяя обратный  оператор $(T_n-1)^{-1}$, мы получаем дискретное уравнение для неизвестной функции $v_{n,m}$:
\eq{\label{quv}(T_m-1)r_{n,m}(v_{n,m})=q_{n,m}\big(p_{n,m+1}^{-1}[(T_n-1)r_{n,m+1}(v_{n,m+1})]),p_{n,m}^{-1}[(T_n-1)r_{n,m}(v_{n,m})]\big).} Дискретное уравнение  \eqref{quv} преобразуется  линеаризуемой заменой \eqref{uv} в уравнение \eqref{our_bN}.

Выбор функции $r_{n,m}$ в замене \eqref{puv} соответствует неавтономному точечному преобразованию неизвестной функции $v_{n,m}$. Из-за интегрирования по $n$, в соотношении \eqref{quv} должна была возникнуть функция интегрирования $\omega_m$. Она убирается за счет выбора $r_{n,m}$.  Кроме того мы выбираем функцию $r_{n,m} $ так, чтобы упростить замену \eqref{uv} и уравнение \eqref{quv}.

Высшие симметрии дискретного уравнения \eqref{quv} в $n$-направлении находятся из \eqref{puv}, см. \cite{gyl16}. Как показывает, например, метод диагонализации $L-A$-пар \cite{mikh15}, функция $p_{n,m}(u_{n,m})$ оказывается плотностью закона сохранения не только дискретного уравнения \eqref{our_bN}, но и ее высших симметрий (\ref{sym_nN1}--\ref{sym_nN3}) с $c=0$.  В силу таких высших симметрий мы имеем
\eq{\label{dthp}D_{\theta_j} p_{n,m}(u_{n,m})=(T_n-1)h^{(j)}_{n,m}(u_{n+j-1,m},u_{n+j-2,m},\ldots, u_{n-j,m}),\quad j=1,2,3.} Дифференцируя \eqref{puv} по $\theta_j$ и интегрируя по $n$, мы находим \eq{ \label{vtheta}r'_{n,m}(v_{n,m})D_{\theta_j} v_{n,m}=h^{(j)}_{n,m}+\widetilde\omega^{(j)}_m,} где все функции вида $u_{n+k,m}$ заменяются функциями $v_{n+k,m}$ в силу \eqref{uv}. Используя условие совместности дискретного уравнения \eqref{quv} и уравнения \eqref{vtheta}, мы уточняем функцию интегрирования $\widetilde\omega^{(j)}_m$. 

Для мастер-симметрий (\ref{sym_nN1}--\ref{sym_nN3}) c $a_n=n$ мы переписываем функцию $D_{\theta_j} p_{n,m}(u_{n,m})$ в терминах $v_{n+k,m}$ при помощи \eqref{uv} и на примерах получаем представление вида: \eq{D_{\theta_j} p_{n,m}(u_{n,m})=(T_n-1)\tilde h^{(j)}_{n,m}(v_{n+j,m},v_{n+j-1,m},\ldots, v_{n-j,m}).} Это представление позволяет найти новую мастер-симметрию так же, как в случае \eqref{dthp}. 

Для получения высших симметрий дискретного уравнения \eqref{quv} в $m$-направлении, мы запишем \eqref{quv} в виде:
\eq{(T_m-1)r_{n,m}(v_{n,m})=q_{n,m}(u_{n,m+1},u_{n,m}).\label{vu_m}}Дифференцируя по $t_j$, $j=1,2$, в силу (\ref{mV1},\ref{sym_mN2}), мы получаем:
\eq{(T_m-1)D_{t_j} r_{n,m}(v_{n,m}) =s^{(j)}_{n,m}(u_{n,m+j+1},u_{n,m+j},\ldots,u_{n,m-j}).} 
Избавляясь в правой части этого соотношения от функций $u_{n,m+k}$ при помощи \eqref{uv}, получаем зависимость от  $v_{n,m+k},v_{n+1,m+k}$. Используя дискретное уравнение \eqref{quv}, приводим это соотношение к виду:\eq{(T_m-1)[ r'_{n,m}(v_{n,m})D_{t_j} v_{n,m}] =\tilde s^{(j)}_{n,m}(v_{n+1,m},v_{n,m+j+1},v_{n,m+j},\ldots,v_{n,m-j}).\label{rt}} В рассматриваемых ниже примерах получается, что зависимость фунции $\tilde s^{(j)}_{n,m}$ от $v_{n+1,m}$ исчезает и эта фунция принадлежит образу оператора $T_m-1$.  Интегрируя по $m$ соотношение \eqref{rt}, получаем дифференциально-разностное уравнение для $v_{n,m}$ с функцией интегрирования $\widehat\omega^{(j)}_n$. Для этого уравнения проверяем не только совместность с дискретным уравнением \eqref{quv}, но и замену \eqref{uv}. При этом уточняется функция $\widehat\omega_n^{(j)}$.

В случае закона сохранения, симметричного \eqref{zsoh}, который имеет вид:
\eq{\label{zsohm}(T_n-1)p_{n,m}(u_{n,m})=(T_m-1)q_{n,m}(u_{n+1,m},u_{n,m}),} где $p'_{n,m}(x)\neq 0$ для любых $n,m$, используется приведенная выше схема с точностью до перестановки $n\leftrightarrow m$.

\subsection{Первая модификация}
Все дискретные уравнения серии \eqref{our_bN} имеют закон сохранения вида \eqref{zsoh}:
\eq{(T_m-1)(\beta^n_N u_{n,m})=(T_n-1)\left[-\frac12\beta_N^n(u_{n,m+1}-1)(u_{n,m}+1)\right].}
Делаем замену переменных
$\beta^n_N u_{n,m}=(T_n-1)(\beta^n_N v_{n,m}),$ которая записывается в виде \eqref{uv} следующим образом:
\eq{u_{n,m}=\beta_N v_{n+1,m}-v_{n,m}\label{uv1}.}  При помощи \eqref{quv} мы получаем дискретное уравнение для $v_{n,m}$:
\eq{(\beta_N\, v_{n+1,m+1}-v_{n,m+1}-1)(\beta_N\, v_{n+1,m}-v_{n,m}+1)+2(v_{n,m+1}-v_{n,m})=0.\label{eqv1}}  

Выпишем для уравнения \eqref{eqv1} четыре высшие симметрии, которые сводятся к (\ref{mV1}--\ref{sym_nN2}) преобразованием \eqref{uv1}. В $m$-направлении мы имеем следующие симметрии для любого $N$:
\eq{\partial_{t_1}v_{n,m}=-2\beta^n_N(v_{n,m+1}-v_{n,m})(v_{n,m}-v_{n,m-1}),\label{vt1}}
\eq{\partial_{t_2}v_{n,m}=4\beta^{2n}_N(v_{n,m+2}-v_{n,m-2})(v_{n,m+1}-v_{n,m})(v_{n,m}-v_{n,m-1}),\label{vt2}} соответствующие \eqref{mV1} и \eqref{sym_mN2}. Связь \eqref{vu_m}, которая используется для построения этих высших симметрий, имеет такую форму: \eq{\label{vu_m2}2(v_{n,m}-v_{n,m+1})=(u_{n,m+1}-1)(u_{n,m}+1).} 

В $n$-направлении при $N=1$, $\beta_N=1$ получаем симметрию первого порядка, соответствующую \eqref{sym_nN1}:
\eq{\partial_{\theta_1}v_{n,m}=-a_n\frac{(v_{n+1,m}-v_{n,m})(v_{n,m}-v_{n-1,m})+1}{v_{n+1,m}-v_{n-1,m}}+cv_{n,m},\label{v_nN1}}
где $a_n=b+cn$ как в \eqref{sym_nN1}. При $N=2$, $\beta_N=-1$ высшая симметрия второго порядка, соответствующая \eqref{sym_nN2}, имеет вид: 
\eqs{\partial_{\theta_2}v_{n,m}=[(v_{n+1,m}+v_{n,m})^2-1]\left(a_{n+1}\frac{v_{n+2,m}+v_{n,m}+2v_{n-1,m}}{V_{n,m}}\right.\nonumber\\-\left.a_{n}\frac{v_{n,m}+2v_{n-1,m}+v_{n-2,m}}{V_{n-1,m}}\right)\label{v_nN2}\\+(b_{n}-b_{n+1})(v_{n+1,m}+v_{n,m})+c(v_{n,m}-v_{n+1,m})\nonumber,\\ V_{n,m}=(v_{n+2,m}+v_{n,m})(v_{n+1,m}+v_{n-1,m})+2(v_{n+2,m}v_{n,m}+v_{n+1,m}v_{n-1,m}+1)\nonumber,} где $a_n=b_n+cn$ как в \eqref{sym_nN2}. 

Уточняя функции интегрирования $\widetilde \omega_m^{(j)},\widehat \omega_n^{(j)}$, возникающие при построении этих высших симметрий, мы во всех случаях приходим к добавке вида $\nu \beta_N^{-n}$ с произвольным постоянным коэффициентом $\nu$, которую мы опускаем. Это сделано по следующей причине: дискретное уравнение \eqref{quv} очевидным образом инвариатно относительно однопараметрической группы преобразований $v_{n,m}\rightarrow v_{n,m}+\tau \beta_N^{-n}$, соответствующей точечной симметрии $\partial_\tau v_{n,m}=\beta_N^{-n}$. Эта точечная симметрия объясняет наличие вышеупомянутой добавки в высших симметриях.

Заметим, что при $N=2$ автономное дискретное уравнение \eqref{eqv1} имеет в обоих направлениях автономные высшие симметрии второго порядка. Симметрия \eqref{vt2} становится автономной, т.к. $\beta_2=-1$, а среди симметрий \eqref{v_nN2} имеется автономный частный случай с $b_n\equiv1,c=0.$

\subsection{О новизне примеров, построенных в этой работе} 
Дискретное уравнение \eqref{eqv1} c $N=1$, т.е. $\beta_N=1$, совпадает с уравнением \cite[(T4), Table 1, p. 17]{ly11} с точностью до замены переменных $v_{n,m}=-2u_{m,n}+m$. Остальные уравнения серии \eqref{eqv1} являются, по-видимому, новыми. 
То же самое верно для остальных модифицированных серий дискретных уравнений. Все уравнения серий являются новыми за исключением случая $N=1$ в некоторых из них.

При каждом фиксированном $n$, высшие симметрии в $m$-направлении \eqref{vt1} и \eqref{vt2} представляют собой одну из известных модификаций уравнения Вольтерры и ее высшую симметрию. Полная классификация интегрируемых уравнений типа Вольтерры была проведена в \cite{y83}, см. подробности в обзоре \cite{y06}. Симметрия \eqref{vt1} соответствует частному случаю уравнения (V6) из списка уравнений типа Вольтерры на страницах R596--R597 \cite{y06}.
Все остальные высшие симметрии в $m$-направлении, выписанные ниже, соответствуют дифференциально-разностным уравнениям аналогичной природы, причем,  все трехточечные примеры принадлежат тому же списку из \cite{y06}.

При каждом фиксированном $m$, высшая симметрия в $n$-направлении \eqref{v_nN1} определяет интегрируемое дифференциально-разностное уравнение при $b=1,c=0$ и его мастер-симметрию при $b=0,c=1$. Это дифференциально-разностное уравнение является частным случаем уравнения (V4) с $\nu=0$ из \cite{y06}. Для всех уравнений вида (V4) с $\nu=0$ в работе \cite{asy00} построены мастер-симметрии с явной зависимостью от своего времени; смотри более подробное изложение в \cite[Section 4.3]{lpsy08}. В частности, в нашем случае мастер-симметрия имеет вид: 
\eq{\partial_{\tau}v_{n}=-\frac{n}{\tau+1}\frac{(v_{n+1}-v_{n})(v_{n}-v_{n-1})+(\tau+1)^2}{v_{n+1}-v_{n-1}},\label{v_ntau}}
где мы для краткости опустили индекс $m$. В отличие от \eqref{v_ntau}, мастер-симметрия, полученная из \eqref{v_nN1}, не имеет зависимости от своего времени и позволяет строить высшие симметрии не только для дифференциально-разностного уравнения, но и для дискретного уравнения \eqref{eqv1} с $N=1$. Кроме того, она возникает естественным образом при построении высших симметрий дискретного уравнения.\footnote{Не вдаваясь в подробности, отметим, что имеется связь между этими мастер-симметриями, т.е. пример, полученный из \eqref{v_nN1}, существенно новым не является.}

В других примерах, высшие симметрии в $n$-направлении, соответствующие $N=1$ в дискретном уравнении, в частном случе $b=1,c=0$ определяют известные дифференциально-разностные уравнения типа Вольтерры из \cite{y06}. Мастер-симметрии, соответствующие случаю $b=0,c=1$, также большей частью известны. 

При фиксированном $m$ высшая симметрия \eqref{v_nN2} определяет интегрируемое пятиточечное автономное дифференциально-разностное уравнение при $b_n\equiv 1,c=0$, его неавтономную симметрию при $b_n= (-1)^n,c=0$ и мастер-симметрию при $b_n\equiv 0,c=1$. Уравнение и мастер-симметрия являются, по-видимому, новыми. 

Остальные высшие симметрии в $n$-направлении, представленные ниже и соответствующие случаям $N=2$ и $N=3$ в дискретных уравнениях, порождают примеры пяти- и семиточечных дифференциально-разностных уравнений. Условие $b_n\equiv1,c=0$ выделяет среди них автономные интегрируемые случаи. Для каждого их них мы автоматически получаем одну или две неавтономные симметрии и мастер-симметрию. Все примеры такого рода, включая неавтономные уравнения и мастер-симметрии, являются по-видимому новыми, за исключением двух отмеченных ниже пятиточечных автономных уравнений.

\subsection{Вторая модификация}
Все дискретные уравнения серии \eqref{our_bN} имеют закон сохранения вида:
\eq{(T_m-1)\log\frac{u_{n,m}+1}{u_{n,m}-1}=(T_n-1)\log\left(\beta_N^n(u_{n,m+1}-1)(u_{n,m}+1)\right).}
Для упрощения нового дискретного уравнения \eqref{quv}, выберем функцию $r_{n,m}$ в соотношении (\ref{puv}) следующим образом: $r_{n,m}=\log w_{n,m}+m\log 4,$ где $w_{n,m}$ -- новая неизвестная функция. 

Тогда из \eqref{puv} получаем:
\eq{\frac{u_{n,m}+1}{u_{n,m}-1}=\frac{w_{n+1,m}}{w_{n,m}},} т.е. в явном виде мы имеем преобразование  \eq{u_{n,m}=\frac{w_{n+1,m}+w_{n,m}}{w_{n+1,m}-w_{n,m}}.\label{uw}} Дискретное уравнение \eqref{quv} на $w_{n,m}$ можно записать следующим образом:
\eq{(w_{n+1,m+1}-w_{n,m+1})(w_{n+1,m}-w_{n,m})=\beta_N^nw_{n+1,m}w_{n,m}.\label{eqw}} Мы получили серию модифицированных дискретных уравнений с одним неавтономным коэффициентом $\beta_N^n.$ Первое уравнение этой серии, соответствующее $N=1$, в переменной $u_{n,m}=(-1)^nw_{m,n}$ совпадает с (T6) из \cite{ly11}.

Связь \eqref{vu_m}, которая используется для построения высших симметрий в $m$-направ\-ле\-нии, имеет вид: \eq{4\frac{w_{n,m+1}}{w_{n,m}}=\beta_N^n(u_{n,m+1}-1)(u_{n,m}+1).\label{wu_m}}
Обе симметрии в этом направлении оказываются автономными:
\eq{\frac1{4}\partial_{t_1}w_{n,m}=w_{n,m+1}+\frac{w_{n,m}^2}{w_{n,m-1}},\label{wt1}}
\eq{\frac1{16}\partial_{t_2}w_{n,m}=w_{n,m+2}+\frac{w_{n,m+1}^2}{w_{n,m}}+\frac{2w_{n,m+1}w_{n,m}}{w_{n,m-1}}+\frac{w_{n,m}^3}{w_{n,m-1}^2}+\frac{w_{n,m}^2}{w_{n,m-2}}.}

В $n$-направлении при $N=1$, $\beta_N=1$ получаем симметрию первого порядка, соответствующую \eqref{sym_nN1}:
\eq{\partial_{\theta_1}w_{n,m}=-a_n\frac{(w_{n+1,m}-w_{n,m})(w_{n,m}-w_{n-1,m})}{w_{n+1,m}-w_{n-1,m}}+cmw_{n,m},\label{wth1}}
где $a_n=b+cn$ как в \eqref{sym_nN1}. Автономный частный случай $b=1,c=0$ этой симметрии является инвариантным относительно дробно-линейных преобразований $w_{n,m}$ (или Мёбиус-инвариантным). Это -- ни что иное, как хорошо известная шварциан-версия 
цепочки Вольтерры.

Удобно ввести обозначения из \cite[Section 4]{a16_1}:\eqn{\label{XY}Y_{n,m}=\frac{(w_{n+1,m}-w_{n,m})(w_{n,m}-w_{n-1,m})}{w_{n+1,m}-w_{n-1,m}},\\
X_{n,m}=\frac{(w_{n+1,m}-w_{n,m})(w_{n-1,m}-w_{n-2,m})}{(w_{n+1,m}-w_{n-1,m})(w_{n,m}-w_{n-2,m})}.} Тогда уравнение \eqref{wth1} легко записывается в терминах $Y_{n,m}.$ 
При $N=2$, $\beta_N=-1$ высшая симметрия второго порядка, соответствующая \eqref{sym_nN2}, имеет вид: 
\eqs{\partial_{\theta_2}w_{n,m}=Y_{n,m}\left(\frac{a_{n+1}}{2X_{n+1,m}-1}+\frac{a_{n}}{2X_{n,m}-1}\right)+2cmw_{n,m},\label{wth2}} где $a_n=b_n+cn$ как в \eqref{sym_nN2}. Очевидно, что при $c=0$ она является Мёбиус-инвариантной относительно $w_{n,m}$ как и \eqref{wth1}. Фиксируя переменную $m$, мы видим, что с точностью до растяжения $\theta_2$, автономный случай $b_n\equiv1,c=0$ совпадает с известным уравнением \cite[(32)]{a16_1}. Для этого уравнения здесь найдены неавтономная симметрия \eqref{wth2} с $b_n=(-1)^n,c=0$ и мастер-симметрия, соответствующая $b_n\equiv 0,c=1$. Кроме того, дискретное уравнение \eqref{eqw} с $N=2$, совместное с симметрией \eqref{wth2}, задает для этого известного уравнения автопреобразование Бэклунда, см. подробности в \cite[Section 4.1]{lpsy08}.

При $N=3$ высшая симметрия третьего порядка, соответствующая \eqref{sym_nN3}, имеет вид:
\eqn{\partial_{\theta_3}w_{n,m}&=Y_{n,m}\left(\frac{a_{n+2}(W_{n+2,m}+\beta_3)}{\beta_3 W_{n+1,m}-W_{n+2,m}}+\frac{a_{n+1}(\beta_3-1)}{\beta_3 W_{n,m}-W_{n+1,m}}-\frac{a_n(\beta_3 W_{n-1,m}+1)}{\beta_3 W_{n-1,m}-W_{n,m}}\right)\\&+3cmw_{n,m},\qquad\qquad W_{n,m}=3X_{n,m}-1,\label{wth3}} где $X_{n,m},Y_{n,m}$ -- функции, определенные в \eqref{XY}, а $\beta_3$ и $a_n=b_n+cn$ определяются как в уравнении \eqref{sym_nN3}. Очевидно, что при фиксированном $m$ и $c=0$ уравнение \eqref{wth3} является Мёбиус-инвариантным. Оно содержит три коммутирующих частных случая, выделенных условиями $b_n\equiv 1,b_n=\beta_3^n$ и $b_n=\beta_3^{2n}$. Первый из них является новым автономным семиточечным Мёбиус-инвариантным интегрируемым уравнением. Случай $b_n\equiv0,c=1$ задает мастер-симметрию для всех трех уравнений. Уравнение \eqref{eqw} с $N=3$ задает для них автопреобразование Бэклунда.

\subsection{Третья модификация}
Здесь мы стартуем с уравнений \eqref{eqw}, которые при любом $N$ имеют закон сохранения
\eq{(T_n-1)\left[(-1)^{n+m}\log w_{n,m}\right]=(T_m-1)\left[(-1)^{n+m}\log\left(\beta_N^{-n/2}(w_{n,m}-w_{n+1,m})\right)\right].}
Этот закон сохранения имеет вид \eqref{zsohm} и формула, симметричная \eqref{puv}, такова:
\eq{(-1)^{n+m}\log w_{n,m}=(T_m-1)\left[(-1)^{n+m+1}\log z_{n,m}\right].} Она приводит к простому линеаризуемому преобразованию
\eq{w_{n,m}=z_{n,m+1}z_{n,m},\label{wz0}} которое позволяет получить следующее дискретное уравнение:
\eq{\label{eqz0}z_{n,m+1}z_{n,m}-z_{n+1,m+1}z_{n+1,m}=\beta_N^{n/2}z_{n+1,m}z_{n,m}.} Соотношение для построения высших симметрий в $n$-направлении имеет вид: \eq{\beta^{n/2}_N z_{n,m}z_{n+1,m}=w_{n,m}-w_{n+1,m}.\label{wzn0}} 

Недостатком уравнения \eqref{eqz0} является наличие полуцелых степеней числа $\beta_N$, что неудобно для исследований. 
Используя неавтономное точечное  преобразование $z_{n,m}^{new}=\beta_{N}^{\frac{n(n-1)}4} z_{n,m}$, мы приводим его к виду, аналогичному \eqref{eqw}: 
\eq{\label{eqz}z_{n,m+1}z_{n,m}-\beta_N^{-n}z_{n+1,m+1}z_{n+1,m}=z_{n+1,m}z_{n,m}.}
Это дискретное уравнение при $\beta_N=1$ не является новым, так как после преобразования $u_{n,m}=i^{n-m}z_{m,n}$ оно совпадает с (T7), $c_2=0$, из \cite{ly11}.

Соотношения \eqref{wz0} и \eqref{wzn0}, необходимые для построения высших симметрий, переписываются так:
\eq{w_{n,m}=\beta_N^{-\frac{n(n-1)}2}z_{n,m+1}z_{n,m},\qquad \beta^{-\frac{n(n-1)}2}_N z_{n,m}z_{n+1,m}=w_{n,m}-w_{n+1,m}.\label{wz2}}
Заметим, что в них также нет полуцелых степений $\beta_N$, так как число $\frac{n(n-1)}2$ всегда целое.

Две высшие симметрии дискретного уравнения \eqref{eqz} в $m$-направлении независимо от номера $N$ имеют вид:
\eq{\frac1{4}\partial_{t_1}z_{n,m}=\frac{z_{n,m+1}z_{n,m}}{z_{n,m-1}},}
\eq{\frac1{16}\partial_{t_2}z_{n,m}=\frac{z_{n,m+2}z_{n,m+1}}{z_{n,m-1}}+\frac{z_{n,m+1}^2z_{n,m}}{z_{n,m-1}^2}+\frac{z_{n,m+1}z_{n,m}^2}{z_{n,m-1}z_{n,m-2}}.}

При $N=1$, $\beta_N=1$ получаем из   \eqref{wth1} высшую симметрию первого порядка в $n$-направлении:
\eq{2\partial_{\theta_1}z_{n,m}=-a_nz_{n,m}\frac{z_{n+1,m}-z_{n-1,m}}{z_{n+1,m}+z_{n-1,m}}+cz_{n,m}\left(m-\frac12\right)   , \label{zth1}}где $a_n=b+cn$ как в \eqref{sym_nN1}. После неавтономного растяжения $z_{n,m}^{new}= i^n z_{n,m}$, это уравнение при любом фиксированном $m$ совпадает с известным уравнением из \cite{y06}, мастер-симметрия которого также известна \cite{asy00,lpsy08}.

При $N=2$, $\beta_N=-1$ высшая симметрия в $n$-направлении второго порядка, полученная из \eqref{wth2}, имеет вид: 
\eqn{\partial_{\theta_2}z_{n,m}=&z_{n,m}\left(\frac{a_n}{1+(-1)^n\frac{z_{n+1,m}}{z_{n-1,m}}\frac{z_{n-2,m}-(-1)^nz_{n,m}}{z_{n-2,m}+(-1)^nz_{n,m}}}\right.\\-&\left.\frac{a_{n+1}}{1+(-1)^n\frac{z_{n-1,m}}{z_{n+1,m}}\frac{z_{n+2,m}-(-1)^nz_{n,m}}{z_{n+2,m}+(-1)^nz_{n,m}}}+cm+\frac{b_{n+1}-b_n}2\right),\label{zth2}} где $a_n=b_n+cn$ как в \eqref{sym_nN2}.
Чтобы избавиться от коэффициента $(-1)^n$, мы используем неавтономное точечное преобразование $z_{n,m}^{new}=i^{-\frac{(n-1)^2}2-\frac14}z_{n,m}$ и получаем:
\eqn{\partial_{\theta_2}z_{n,m}=z_{n,m}\left(\frac{a_n}{1-\frac{z_{n+1,m}}{z_{n-1,m}}\frac{z_{n-2,m}-z_{n,m}}{z_{n-2,m}+z_{n,m}}}-\frac{a_{n+1}}{1-\frac{z_{n-1,m}}{z_{n+1,m}}\frac{z_{n+2,m}-z_{n,m}}{z_{n+2,m}+z_{n,m}}}+cm+\frac{b_{n+1}-b_n}2\right).\label{zth2a}} Так как здесь $c$ -- постоянная, а $b_n$ -- произвольная двухпериодическая функция, то при каждом фиксированном $m$ мы имеем три случая: автономное дифференциально-разностное уравнение при $a_n\equiv 1$, его симметрию при $a_n=(-1)^n$ и их общую  мастер-симметрию при $a_n=n$.

В этом случае второе соотношение из \eqref{wz2} превращается в \eq{z_{n,m}z_{n+1,m}=w_{n,m}-w_{n+1,m}\label{wz3},} и при фиксированном $m$ оно связывает дифференциально-разностные уравнения \eqref{wth2} и \eqref{zth2a}. При $a_n\equiv 1$ уравнение \eqref{zth2a} становится автономным и оно связано автономным преобразованием \eqref{wz3} с известным автономным уравнением \eqref{wth2} с $a_n\equiv1$. Преобразование \eqref{wz3} является линеаризуемым, но неявным в обоих направлениях \cite{gyl16}.

\subsection{Четвертая модификация}

Правую и левую часть соотношения \eqref{vu_m2} можно обозначить через $y_{n,m}$, введя таким образом новую неизвестную функцию. 
Как следует из \eqref{mV1} и \eqref{vt1}, она удовлетворяет уравнению \eq{\partial_{t_1} y_{n,m}=\beta_N^n y_{n,m}(y_{n,m+1}-y_{n,m-1}),\label{V1}}которое является уравнением Вольтерры при любом фиксированном $n$. Это естественно, так как одно из заданных здесь преобразований \eq{y_{n,m}=(u_{n,m+1}-1)(u_{n,m}+1)\label{miura}} есть ни что иное, как известное преобразование модифицированного уравнения Вольтерры  в уравнение Вольтерры. Другое преобразование $$y_{n,m}=2(v_{n,m}-v_{n,m+1}),$$ связывающее \eqref{vt1} с \eqref{V1}, известно так же, как и \eqref{miura}, см. например обзор \cite{y06}. 
Обозначая в соотношении \eqref{wu_m}  правую и левую часть через $\beta_N^n y_{n,m}$, мы снова получаем связи между \eqref{wt1} и \eqref{mV1} с тем же самым уравнением \eqref{V1}.

Подобного результата можно ожидать от соотношения \eqref{wz3}. Здесь мы покажем, что преобразование \eq{y_{n,m}=w_{n,m}-w_{n+1,m},\label{yw}} полученное из \eqref{wz3}, позволает вывести из дискретного уравнения \eqref{eqw} еще одну модификацию вместе с высшими симметриями. Заметим, что в отличие от (\ref{uv1},\ref{uw},\ref{wz0}) преобразование \eqref{yw} имеет другое направление в том смысле, что позволяет выразить явным образом новую неизвестную функцию $y_{n,m}$ через старую $w_{n,m}$.

Преобразование \eqref{yw} позволяет без труда переписывать высшие симметрии в $n$-направлении. Симметрии \eqref{wth1} и \eqref{wth2} принимают вид:
\eq{\partial_{\theta_1}y_{n,m}=y_{n,m}^2\left(\frac{a_{n+1}}{y_{n+1,m}+y_{n,m}}-\frac{a_{n}}{y_{n,m}+y_{n-1,m}}\right)+c(m-1)y_{n,m},}
\eqn{\partial_{\theta_2}y_{n,m}=&-y_{n,m}^2\left(\frac{a_{n+2}(y_{n+2,m}-y_{n+1,m})}{\Upsilon_{n+1,m}}+\frac{a_{n+1}(y_{n+1,m}-y_{n-1,m})}{\Upsilon_{n,m}}\right.\\ &\left.+\frac{a_{n}(y_{n-1,m}-y_{n-2,m})}{\Upsilon_{n-1,m}}\right)+2c(m-1)y_{n,m},\\ &\Upsilon_{n,m}=(y_{n+1,m}+y_{n,m})(y_{n,m}+y_{n-1,m})-2y_{n+1,m}y_{n-1,m},} где функции $a_n$ -- такие же, как в \eqref{sym_nN1} и \eqref{sym_nN2}.
При желании можно переписать и высшую симметрию \eqref{wth3}.

Новое дискретное уравнение для функции $y_{n,m}$ и его высшие симметрии в $m$-направлении будут содержать квадратные корни. Однако, нам важен тот факт, что это преобразование \eqref{yw} позволяет их построить. 

Перепишем дискретное уравнение \eqref{eqw} при помощи преобразования \eqref{yw} в виде квадратного уравнения для неизвестной функции $w_{n,m}$:
\eq{\label{kv_eq}y_{n,m+1}y_{n,m}=\beta_N^{n}(w_{n,m}-y_{n,m})w_{n,m}.} Обозначим
\eq{\Theta_{n,m}=1+4\beta^{-n}_N\frac{y_{n,m+1}}{y_{n,m}},} тогда решение \eqref{kv_eq} запишется в виде
\eq{\label{wy}w_{n,m}=\frac{y_{n,m}}2\left(1+\sqrt{\Theta_{n,m}}\right),} где $\sqrt{\Theta_{n,m}}$ -- любая из ветвей функции квадратного корня. Применяя к соотношению \eqref{wy} оператор $1-T_n$, получаем новое дискретное уравнение
\eq{y_{n+1,m}\left(1+\sqrt{\Theta_{n+1,m}}\right)+y_{n,m}\left(1-\sqrt{\Theta_{n,m}}\right)=0.}

Для нахождения простейшей высшей симметрии в $m$-направлении, продифференцируем преобразование \eqref{yw} по $t_1$ в силу  симметрии \eqref{wt1}. Затем в правой части полученного выражения исключим функции $w_{n+1,m+1}$, $w_{n+1,m}$, $w_{n+1,m-1}$ в силу преобразования \eqref{yw}, а потом $w_{n,m},w_{n,m-1}$ -- в силу \eqref{wy}. Получим высшую симметрию 
\eq{\partial _{t_1}y_{n,m}=2\beta_N^ny_{n,m}\left(\sqrt {\Theta_{n,m}}\sqrt{\Theta_{n,m-1}}-1\right).}
Точечное преобразование $y_{n,m}=\exp\hat y_{n,m}$ позволяет получить в терминах $\hat y_{n,m}$ симметрию, которая при каждом фиксированном $n$ имеет вид известного уравнения \cite[Page R596, (V6)]{y06}.

\section{Преобразования, обратимые на решениях дискретного уравнения}\label{secobr}
В этой секции мы строим для дискретных уравнений неточечные преобразования, обратимые на решениях этих дискретных уравнений. Для высших симметрий дискретных уравнений такие преобразования порождают преобразования типа Миуры или сложно-линеаризуемые преобразования \cite{gyl16}. При этом мы используем метод, разработанный в \cite{s10,y91} для дискретных и полудискретных уравнений. Другая версия того же метода, предназначенная для построения преобразований типа Миуры, представлена в \cite{y93,y94}.

В результате мы получаем еще две модификации серии \eqref{our_bN} дискретных уравнений вместе с их высшими симметриями.

\subsection{Первая модификация}\label{sec_first}
Перепишем дискретное уравнение \eqref{our_bN} в виде:
\eq{\frac{u_{n,m}-1}{u_{n+1,m}+1}=\beta_N\frac{u_{n+1,m+1}-1}{u_{n,m+1}+1}.}
Это представление позволяет ввести новую неизвестную функцию $v_{n,m}$ следующим образом:
\eq{v_{n,m+1}=\frac{u_{n,m}-1}{u_{n+1,m}+1},\quad v_{n,m}=\beta_N\frac{u_{n+1,m}-1}{u_{n,m}+1}.\label{vu2}} Полученные выражения можно разрешить относительно старых неизвестных $u_{n,m},$ $u_{n+1,m}$:
\eq{u_{n,m}=\frac{v_{n,m+1}v_{n,m}+2\beta_Nv_{n,m+1}+\beta_N}{\beta_N-v_{n,m+1}v_{n,m}},\quad u_{n+1,m}=\frac{v_{n,m+1}v_{n,m}+2v_{n,m}+\beta_N}{\beta_N-v_{n,m+1}v_{n,m}}.\label{uv2}}

Записывая последние выражения в одной точке $u_{n+1,m}$, получаем новое дискретное уравнение. Оно существенно упрощается, если его переписать в терминах $(u_{n+1,m}-1)/(u_{n+1,m}+1)$:
\seq{\frac{v_{n,m}(v_{n,m+1}+1)}{v_{n,m}+\beta_N}=\frac{v_{n+1,m+1}(v_{n+1,m}+\beta_N)}{\beta_N(v_{n+1,m+1}+1)}.}
Еще одна эквивалентная форма записи имеет вид:
\eq{(v_{n+1,m}+\bN)(1+\beta_N v_{n,m}^{-1})=\bN (v_{n,m+1}+1)(1+v_{n+1,m+1}^{-1}).\label{eqv2}}

Мы видим, что модифицированное уравнение \eqref{eqv2} получается из \eqref{our_bN} при помощи преобразования \eqref{vu2}, которое обратимо на решениях дискретного уравнения \eqref{our_bN}. Обратное преобразование имеет вид \eqref{uv2}.
Заметим, что при $\bN=1$ ничего нового не получается, так как точечное преобразование\eq{\hat v_{n,m}=\frac{1-v_{n,m}}{1+v_{n,m}}\label{hatv}} приводит  к дискретному уравнению, которое после замены переменых $n\leftrightarrow m$ становится уравнением \eqref{our_bN} c $\bN=1.$  

Обратимое преобразование \eqref{vu2} позволяет переписывать регулярным образом высшие симметрии \cite{y91}. Так из симметрий \eqref{mV1} и \eqref{sym_mN2} уравнения \eqref{our_bN} мы получаем две простейшие высшие симметрии в $m$-направлении для нового дискретного уравнения \eqref{eqv2}:
\eqn{\partial_{t_1}v_{n,m}=4\beta^{n+1}_Nv_{n,m}(v_{n,m}+1)(v_{n,m}+\beta_N)\frac{v_{n,m+1}-v_{n,m-1}}{V_{n,m+1}V_{n,m}},\label{v2t1}
\\V_{n,m}={v_{n,m}v_{n,m-1}-\beta_N},}
\eqn{\partial_{t_2}v_{n,m}=16\beta^{2n+2}_N(v_{n,m}+1)(v_{n,m}+\beta_N)\left[-\frac{v_{n,m}(v_{n,m+1}+1)(v_{n,m+1}+\beta_N)}{V_{n,m+2}V_{n,m+1}^2}\right.\\+\left.\frac{v_{n,m}(v_{n,m-1}+1)(v_{n,m-1}+\beta_N)}{V_{n,m}^2V_{n,m-1}}-\frac{v_{n,m+1}+v_{n,m}+\beta_N+1}{V_{n,m+1}^2}\right.\\\left.+\frac{v_{n,m}+v_{n,m-1}+\beta_N+1}{V_{n,m}^2}+\frac{\beta_N(v_{n,m}+1)(v_{n,m}+\beta_N)(v_{n,m+1}-v_{n,m-1})}{V_{n,m+1}^2V_{n,m}^2}\right].\label{v2t2}} Отметим, что точечное преобразование $v^{new}_{n,m}=\frac{\sqrt{\beta_N}-v_{n,m}}{\sqrt{\beta_N}+v_{n,m}}$ переводит симметрию \eqref{v2t1} в известное уравнение вида \cite[(V2)]{y06} при любом фиксированном $n$.

Перейдем к симметриям в $n$-направлении и рассмотрим сначала случай $N=1$. Высшая симметрия, полученная из \eqref{sym_nN1} и затем переписанная в терминах $\hat v_{n,m}$ из \eqref{hatv}, принимает при любом фиксированном $m$ вид модифицированного уравнения Вольтерры с известной мастер-симметрией \cite{ztof91}:
\eq{4\partial_{\theta_1}\hat v_{n,m}=(\hat v_{n,m}^2-1)(a_{n+2}\hat v_{n+1,m}-a_n \hat v_{n-1,m}),\quad a_n=b+cn.\label{mVolt}}
Простейшая высшая симметрия при $N=2$, полученная из \eqref{sym_nN2}, имеет вид: 
\eqn{\partial_{\theta_2}v_{n,m}=v_{n,m}(1-T_n^2)\left(a_{n+1}\frac{v_{n,m}-1}{\widehat V_{n,m}}-a_n\frac{v_{n-2,m}+1}{\widehat V_{n-1,m}}\right)-2cv_{n,m},\\ \widehat V_{n,m}=v_{n,m}v_{n-1,m}-v_{n,m}+v_{n-1,m}+1,\label{v2th2}}где $a_n$ определяется как в \eqref{sym_nN2}. Для каждого фиксированного $m$, она порождает при $a_n\equiv 1$ пятиточечный аналог модифицированного уравнения Вольтерры. С другой стороны, этот случай  $a_n\equiv1$ является новым примером автономного дифференциаль\-но-разностного уравнения с неавтономной симметрией $a_n=(-1)^n$ и мастер-симмет\-ри\-ей $a_n=n$. 

Каждое из преобразований \eqref{vu2} преобразует высшую симметрию \eqref{sym_nN2} в \eqref{v2th2} напрямую, поскольку они полностью определены на прямой $n$ так же, как и симметрии. 
В этом их отличие от проебразований \eqref{uv2}, которые связывают те же симметрии только на решениях дискретного уравнения \eqref{eqv2}.  
В отличие от преобразований, рассмотренных в секции \ref{seclin}, преобразования \eqref{vu2} линеаризуются более сложным образом. Например, первое из них является композицией преобразований:
\eq{v_{n,m+1}=\frac{y_{n+1,m}}{y_{n,m}},\quad y_{n,m}=z_{n,m}-z_{n-1,m},\quad u_{n,m}=\frac{z_{n-1,m}+z_{n,m}}{z_{n-1,m}-z_{n,m}}. }Они являются линеаризуемыми, т.к. с точностью до сдвига $T_n$ легко записываются в виде \eqref{lin_tr}.

Теми же самыми сложно-линеаризуемыми преобразованиями \eqref{vu2} высшая симметрия \eqref{sym_nN3} связана с
\eqn{\partial_{\theta_3}v_{n,m}=v_{n,m}(T_n^2-1)\left[{a_{n+2}}\beta_3\frac{v_{n+1,m}v_{n,m}+\beta_3 v_{n+1,m}+v_{n,m}+\beta_3^2}{\widetilde V_{n,m}}\right.\\+a_{n+1}\beta^2_3\frac{(v_{n,m}+\beta_3)(v_{n-2,m}+1)}{\widetilde V_{n-1,m}}\\\left.+a_n\frac{v_{n-2,m}v_{n-3,m}+\beta_3v_{n-2,m}+v_{n-3,m}+1}{\widetilde V_{n-2,m}}\right]-3cv_{n,m},\\\widetilde V_{n,m}=(v_{n+1,m}+1)(v_{n,m}+\beta_3^2)(v_{n-1,m}+\beta_3)+(\beta_3+2)(v_{n-1,m}-v_{n+1,m}),\label{v2th3}}где коэффициенты $a_n,c,\beta_3$ -- такие же, как в \eqref{sym_nN3}. Здесь при каждом фиксированном $m$ содержится один автономный семиточечный аналог модифицированного уравнения Вольтерры, две его неавтономные коммутирующие симметрии и одна мастер-симметрия.

Так же как \eqref{eqv1}, уравнение \eqref{eqv2} является автономной модификацией автономного дискретного уравнения \eqref{our_bN}. Как и в случае \eqref{our_bN} можно показать, что  оно является примером автономного дискретного уравнения с двумя иерархиями автономных высших симметрий. Например, при $N=2$ и $\bN=-1$, простейшими автономными высшими симметриями являются \eqref{v2t2} и \eqref{v2th2} c $a_n\equiv1$, которые оказываются пятиточечными.

\subsection{Вторая модификация}\label{sec_sec}
Обозначим правую и левую часть дискретного уравнения \eqref{eqv2} через $3\beta_N/w_{n,m+1}$ и из этих двух равенств получаем формулы:
\eq{w_{n,m}=\frac{3v_{n+1,m}}{(v_{n+1,m}+1)(v_{n,m}+1)},\quad w_{n,m+1}=\frac{3\beta_Nv_{n,m}}{(v_{n+1,m}+\beta_N)(v_{n,m}+\beta_N)}.\label{wv2}}
Они определяют преобразование $v_{n,m},v_{n+1,m}$ в $w_{n,m},w_{n,m+1}$, обратимое на решениях дискретного уравнения \eqref{eqv2}. Как и в случае \eqref{vu2}, мы можем построить в терминах $w_{n,m}$ новое дискретное уравнение и его высшие симметрии по обоим направлениям. Однако, обратное преобразование содержит квадратные корни, поэтому дискретное уравнение и симметрии в $m$-направлении также будут содержать квадратные корни.  
Формулы \eqref{wv2} обеспечивают явные рациональные преобразования для симметрий в $n$-направлении, которые также оказываются рациональными. Мы выпишем в основном такие симметрии, а соответствующее дискретное уравнение покажем только в одном важном случае.

В отличие от \eqref{vu2} мы имеем здесь не линеаризуемые преобразования, а преобразования типа Миуры. Действительно, в терминах $\hat v_{n,m}$ из \eqref{hatv} и \eq{\check v_{n,m}=\frac{\beta_N-v_{n,m}}{\beta_N+v_{n,m}},\label{chv}} мы имеем
\eq{w_{n,m}=-\frac34(\hat v_{n+1,m}-1)(\hat v_{n,m}+1),\quad w_{n,m+1}=-\frac 34(\check v_{n+1,m}+1)(\check v_{n,m}-1).\label{wvhat}} Это известные дискретные преобразования Миуры, связывающие уравнение Вольтерры и его известную модификацию, и они получены здесь при помощи точечных преобразований функции $v_{n,m}.$  

В случае $N=1$, $\beta_N=1$, используя первое из преобразований \eqref{wvhat}, мы получаем из симметрии \eqref{mVolt}, записанной в терминах $\hat v_{n,m}$, следующую высшую симметрию в $n$-направлении:
\eq{-3\partial_{\theta_1}w_{n,m}=w_{n,m}(a_{n+3}w_{n+1,m}-a_nw_{n-1,m}+cw_{n,m}-3c),\quad a_n=b+cn.\label{Volt}}
Способ переписывания дифференциально-разностных уравнений при помощи таких преобразований объясняется в \cite[Appendix A.2]{gyl17}. При любом фиксированном $m$, симметрия \eqref{Volt} представляет собой уравнение Вольтерры с его известной мастер-симмет\-ри\-ей \cite{ozf89}. 

Если $N=2$, т.е. $\beta_N=-1,$ то мы получаем из симметрии \eqref{v2th2} при помощи первого из преобразований \eqref{wv2} следующую высшую симметрию:
\eq{\partial_{\theta_2}w_{n,m}=w_{n,m}(T_n+1)\left[\frac{a_{n+3}w_{n,m}}{2w_{n+1,m}-3}-\frac{a_nw_{n-1,m}}{2w_{n-2,m}-3}+\frac12(T_n-1)\left(a_n-\frac{3a_{n+1}}{2w_{n-1,m}-3}\right)\right],\label{w2th2}}
где коэффициент $a_n$ определяется как в \eqref{sym_nN2}.

Мы можем получить известное уравнение в терминах
\eq{\wt_{n,m}=-\frac{3}{2w_{n,m}-3}.} Симметрия \eqref{w2th2} записывается следующим образом:
\eqn{-2\partial_{\theta_2}\wt_{n,m}=(\wt_{n,m}-1)\left[a_{n+4}\frac{\wt_{n+2,m}(\wt_{n+1,m}-1)\wt_{n,m}}{\wt_{n+1,m}}-a_{n}\frac{\wt_{n,m}(\wt_{n-1,m}-1)\wt_{n-2,m}}{\wt_{n-1,m}}\right.\\-\left.a_{n+3}\wt_{n+1,m}+a_{n+1}\wt_{n-1,m}+(a_n-a_{n+2})\wt_{n,m}\right].\label{wtth2}}
В частном случае $a_n\equiv1$, при любом фиксированном $m$, она является пятиточечным аналогом уравнения Вольтерры. Это уравнение  ранее было найдено в \cite{a16_1}; уравнение \cite[(39)]{a16_1} совпадает с нашим с точностью до растяжения $\theta_2$ и $\wt_{n,m}$). В статье \cite{a18} отмечено, что оно играет ключевую роль для специального класса дифференциально-разностных уравнений.

Формула \eqref{wtth2} с тем же фиксированным $m$ дает для этого уравнения неавтономную симметрию при $a_n=(-1)^n$ и мастер-симметрию при $a_n=n$. Кроме того, мы можем выписать для этого известного уравнения соответствующее дискретное уравнение, задающее автопреобразование Бэклунда:
\eqn{&\frac{\wt_{n+1,m+1}-1}{\wt_{n+1,m+1}(\wt_{n+1,m}+1)}(\Theta_{n+1,m}-\wt_{n+1,m+1}+\wt_{n+1,m}+2)\\&+\frac{\wt_{n+1,m}-1}{\wt_{n+1,m}(\wt_{n,m+1}-1)}(\Theta_{n,m}+\wt_{n,m+1}-\wt_{n,m}-2)=4,\\ &\Theta_{n,m}=\sqrt{4\wt_{n,m+1}\wt_{n,m}+(\wt_{n,m+1}+\wt_{n,m}+2)^2}.}

При $N=3$ мы можем построить семиточечный аналог уравнения Вольтерры. Используя первое из преобразований \eqref{wv2}, мы получаем из \eqref{v2th3} высшую симметрию
\eqn{3\partial_{\theta_3}w_{n,m}=&w_{n,m}(1+T_n)\left[A_{n,m}+(1-T_n)(w_{n-1,m}B_{n,m})\right],\\
A_{n,m}=&a_{n+4}\frac{\beta_3w_{n+1,m}w_{n,m}}{W_{n+2,m}}-a_{n+2}\frac{(\beta_3+1)w_{n,m}w_{n-1,m}}{W_{n,m}}\\&+a_n\frac{w_{n-1,m}w_{n-2,m}}{W_{n-2,m}}-3c\left(w_{n,m}-\frac32\right),\\B_{n,m}=&a_{n+2}\frac{2\beta_3+1}{W_{n,m}}-a_{n+1}\frac{\beta_3+2}{W_{n-1,m}}+a_n,\\ W_{n,m}=&\beta_3w_{n-1,m}-w_{n,m}+1-\beta_3,\label{w2th3}} 
где коэффициенты $a_n,c,\beta_3$ определяются как в \eqref{sym_nN3}. Если зафиксировать $m$, то мы получаем из симметрии \eqref{w2th3} вышеуказанный аналог при $a_n\equiv1$, его неавтономные коммутирующие симметрии при $a_n=\beta_3^n$ и $a_n=\beta_3^{2n}$, а также мастер-симметрию при $a_n=n$.

\end{document}